\documentclass[showpacs,amsmath,amssymb,twocolumn,pra,superscriptaddress,notitlepage]{revtex4-1}

\usepackage{qcircuit}
\usepackage[dvips]{graphicx}
\usepackage{amsmath,amssymb,amsthm,mathrsfs,amsfonts,dsfont}
\usepackage{subfigure, epsfig}
\usepackage{braket}
\usepackage{bm}
\usepackage{enumerate}
\usepackage{color}
\usepackage{graphicx}

\newcommand{\tr}{\mathrm{Tr}}

\begin{document}


\title{Mitigating algorithmic errors in Hamiltonian simulation}


\author{Suguru Endo }
\affiliation{Department of Materials, University of Oxford, Parks Road, Oxford OX1 3PH, United Kingdom}

\author{Qi Zhao}
\affiliation{Center for Quantum Information, Institute for Interdisciplinary Information Sciences, Tsinghua University, Beijing 100084, China}

\author{Ying Li}
\affiliation{Graduate School of China Academy of Engineering Physics, Beijing 100193, China}

\author{Simon Benjamin}
\affiliation{Department of Materials, University of Oxford, Parks Road, Oxford OX1 3PH, United Kingdom}

\author{Xiao Yuan}
\email{xiao.yuan.ph@gmail.com}
\affiliation{Department of Materials, University of Oxford, Parks Road, Oxford OX1 3PH, United Kingdom}


\begin{abstract}
Quantum computers can efficiently simulate many-body systems.
As a widely used Hamiltonian simulation tool, the Trotter-Suzuki scheme splits the evolution into the number of Trotter steps $N$ and approximates the evolution of each step by a product of exponentials of each individual term of the total Hamiltonian.
The algorithmic error due to the approximation can be reduced by increasing $N$, which however requires a longer circuit and hence inevitably introduces more physical errors.
In this work, we first study such a trade-off and numerically find the optimal number of Trotter steps $N_{\textrm{opt}}$ given a physical error model in a near-term quantum hardware.
Practically, physical errors can be suppressed using recently proposed error mitigation methods.
We then extend physical error mitigation methods to suppress the algorithmic error in Hamiltonian simulation.
By exploiting the simulation results with different numbers of Trotter steps $N\le N_{\textrm{opt}}$, we can infer the exact simulation result within a higher accuracy and hence mitigate algorithmic errors.
We numerically test our scheme with a five qubit system and show significant improvements in the simulation accuracy by applying both physical and algorithmic error mitigations.



\end{abstract}

\maketitle

\section{Introduction}
It is hard to simulate quantum systems using a classical computers, as the computational cost increases exponentially with the system size.
Such a problem can be resolved by quantum simulation, as proposed by Feynman in 1982 \cite{Feynman1982}, saying ``let the computer itself be built of quantum mechanical elements which obey quantum mechanical laws''.
Quantum simulation of many-body systems has become one of the most promising applications of quantum computing.
Given the Hamiltonian $H$ of a system, a vital step is to realise the time evolution operator $U(t)=e^{-iHt}$, which can be used for studying both its dynamic \cite{sethuniversal} and static \cite{aspuru2005simulated} properties.

Several methods have been proposed to efficiently approximate the time evolution operator $U(t)$ \cite{suzuki1991general,childs2018faster,TaylorSeries,low2016hamiltonian,childs2017toward}. Although the latest methods \cite{TaylorSeries,low2016hamiltonian,childs2017toward} have been significantly improved, the simulation accuracy is still limited by finite resources, such as short circuit depth, finite system runtime, and large physical errors in the system.
To study such a limitation, we focus on the the Trotterization method \cite{trottersuzuki}, introduced for quantum simulation by Lloyd \cite{sethuniversal}. Suppose the system Hamiltonian $H$ can be decomposed into a sum of Hamiltonians, $H_k$, that only involve few-body interactions, i.e.~$H=\sum_k H_k$. Then, the Trotterization method approximates the time evolution unitary operator $U(t)=e^{-i\sum_k H_kt}$ by decomposing it into a product form,
\begin{align}
U(t) = \left( \prod_k e^{-i H_k t/N} \right) ^N + O(t^2/N).
\label{trotter}
\end{align}
Here, $N$ is the number of Trotter steps and $O(t^2/N)$ is the algorithmic error due to a finite value of $N$. As $H_k$ only has local interactions, each term $e^{-i H_k t/N}$ can be efficiently realised on a quantum computer. However, as $H_k$ generally does not commute with each other, Trotterization only approximates the  time evolution operator $U(t)$.
By increasing the number of Trotter steps $N$, the algorithmic error $O(t^2/N)$ can be arbitrarily suppressed. However, the circuit depth increases linearly with the number of Trotter steps. A deep circuit introduces more physical errors, which corrupt quantum simulations in noisy intermediate-scale devices \cite{preskill2018quantum}. Consequently, we can only use a small number of Trotter steps for systems without quantum error correction \cite{george}. 

Recently, several error mitigation methods have been introduced to suppress physical errors in shallow circuits \cite{Li2017,PhysRevLett.119.180509, endo2017practical, recoveringnoisefree, samerrormitigation,bonet2018low}. One of the methods relies on extrapolation \cite{Li2017,PhysRevLett.119.180509,endo2017practical}, which works by deliberately increasing the error rate of the quantum hardware, and using the expectation values of several points with higher error rates to infer the error-free value.
Although the error extrapolation method can suppress physical errors in shallow circuits, it fails to work when the circuit length is too long, such that several errors are expected in the circuit.
As such, even with error mitigation, we cannot choose too large a number of Trotter steps. As a result, even if physical errors can be mitigated, the accuracy of the simulations will still be limited by algorithmic errors.

In our work, we study algorithmic and physical errors resulting from implementing Trotterization on near-term devices.
Following~Ref. \cite{george}, we first find the optimal number of Trotter steps $N_{\textrm{opt}}$ for error prone quantum simulations in Sec.~\ref{Sec:optimalN}. In Sec.~\ref{Sec:errormitigation}, we review recently proposed error mitigation methods, and show how physical errors can be suppressed.
We then extend the physical error extrapolation method to suppress the algorithmic error in Trotterization in Sec.~\ref{Sec:algorithmic}. We use expectation values obtained from different number of Trotter steps $N_1$ and $N_2$ $(N_{\textrm{opt}}\ge N_1, N_2)$ and extrapolate the results to estimate a more accurate expectation value than using only the optimal number of Trotter steps $N_{\textrm{opt}}$. 
In Sec.~\ref{Sec:numerics}, we numerically test our result by considering a five qubit Hamiltonian. We numerically show the optimal number of Trotter steps $N_{\textrm{opt}}$ under an inhomogeneous Pauli error model, and apply both physical and algorithmic error mitigations to significantly increase the simulation accuracy. In Sec.~\ref{Sec:discussion}, we conclude our work and discuss its possible extension in general quantum information processing.

\section{The Optimal number of Trotter steps for noisy quantum simulation}\label{Sec:optimalN}
In this section, we first review the theoretical analysis \cite{george} about the optimal number of Trotter steps in Hamiltonian simulation with physical noise.
Denote the channel of the Trotter decomposition $e^{- i H_k t/N}$ as $\mathcal{V}_k$ and the physical noise as an extra channel  $\mathcal{E}_k$, the noisy stroboscopic channel for the $i^{\textrm{th}}$ Trotter step is
\begin{equation}
	\mathcal{E}^{\mathrm{strobo}}_i=\mathcal{E}_L \circ \mathcal{V}_L \circ \mathcal{E}_{L-1} \circ  \mathcal{V}_{L-1}...\circ \mathcal{E}_1 \circ \mathcal{V}_1,
\end{equation}
where $L$ is the number of the local Hamiltonians.
With the number of Trotter steps $N$, the entire noisy channel of the Trotter decomposition is
\begin{align}
\mathcal {E}^{\mathrm{noisyTrotter}}&=  \mathcal{E}^{\mathrm{strobo}}_N \circ  \mathcal{E}^{\mathrm{strobo}}_{N-1} \circ \dots \circ\mathcal{E}^{\mathrm{strobo}}_1.
\end{align}
 The distance between two channels $\mathcal{E}_1$ and $\mathcal{E}_2$ is defined by the trace distance or distinguishability between the output states of the two channels \cite{distinguishability},
\begin{equation}
	D(\mathcal{E}_1,\mathcal{E}_2) = \max_{\rho}\|\mathcal{E}_1(\rho)-\mathcal{E}_2(\rho)\|,
\end{equation}
where the maximisation is over a properly chosen state set and $\|M\| = \tr[\sqrt{M^\dag M}]$ for matrix $M$.
The distance between the ideal channel $\mathcal{U}^{\mathrm{ideal}}$ for the evolution $e^{-iHt}$ and the noisy implementation $\mathcal {E}^{\mathrm{noisyTrotter}}$ is
\begin{equation}\label{ineaq}
	\begin{aligned}
&D(\mathcal{U}^{\mathrm{ideal}},\mathcal {E}^{\mathrm{noisyTrotter}})\\
&\leq \sum_{i=1}^N D(\sqrt [N]{\mathcal{U}^{\mathrm{ideal}}}, \mathcal{E}^{\mathrm{strobo}}_i), \\
& \leq \sum_{i=1}^N D(\sqrt [N]{\mathcal{U}^{\mathrm{ideal}}}, \sqrt [N]{\mathcal{V}^{\mathrm{nonoise}}}) +D(\mathcal{E}^{\mathrm{strobo}}_i, \sqrt [N]{\mathcal{V}^{\mathrm{nonoise}}} ), \\
\end{aligned}
\end{equation}
where the second line follows from the chaining property $D(\mathcal{E}_1 \circ \mathcal{E}_2,  \mathcal{E}_1^\prime \circ \mathcal{E}_2 ^\prime) \leq D(\mathcal{E}_1, \mathcal{E}_1^\prime) + D(\mathcal{E}_2, \mathcal{E}_2^\prime)$, the third line follows from the triangle inequality $D(\mathcal{E}_1 , \mathcal{E}_2) \leq D(\mathcal{E}_1,  \mathcal{E}_3)+D(\mathcal{E}_2, \mathcal{E}_3)$, and $\sqrt [N]{\mathcal{V}^{\mathrm{nonoise}}}$ is the channel of the noise free stroboscopic evolution $\prod_k e^{-i H_k t/N}$ of each Trotter decomposition. Now, we define the algorithmic and physical errors $\varepsilon_{alg}$ and $\varepsilon_{phys}$ as
\begin{equation}\label{ineaq}
	\begin{aligned}
\varepsilon_{alg} & = \sum_{i=1}^N D(\sqrt [N]{\mathcal{U}^{\mathrm{ideal}}}, \sqrt [N]{\mathcal{V}^{\mathrm{nonoise}}})=\frac{\alpha}{N},\\
\varepsilon_{phys} & = \sum_{i=1}^N  D(\mathcal{E}^{\mathrm{strobo}}_i, \sqrt [N]{\mathcal{V}^{\mathrm{nonoise}}} )
= \beta N,
\end{aligned}
\end{equation}
where $\alpha=D(\sqrt [N]{\mathcal{U}^{\mathrm{ideal}}}, \sqrt [N]{\mathcal{V}^{\mathrm{nonoise}}})N^2 $ and $\beta = D(\mathcal{E}^{\mathrm{strobo}}, \sqrt [N]{\mathcal{V}^{\mathrm{nonoise}}} ) $. Here, for simplicity, we assume that the noise model is the same for each stroboscopic sequence, that is $\mathcal{E}^{\mathrm{strobo}}_i$ is the same for different $i$.
Note that $\varepsilon_{alg}\propto1/N$ is because the algorithmic error can be linearly suppressed with an increasing number of Trotters; while $\varepsilon_{phys}\propto N$ is because errors accumulate with a larger number of Trotter steps and hence longer circuits.
By optimising the distance
\begin{equation}
	D(\mathcal{U}^{\mathrm{ideal}},\mathcal {E}^{\mathrm{noisyTrotter}}) = \frac{\alpha}{N}+\beta N,
\end{equation}
we can get the optimised number of Trotter step as
\begin{align}\label{opttrotter}
N_{\mathrm{opt}}=\sqrt{\alpha/\beta},
\end{align}
with the corresponding trace distance $D= 2 \sqrt{\alpha \beta }$.
Therefore, due to the existence of physical errors,  we cannot choose an infinitely large number of Trotter steps to suppress the algorithmic error. Although it is not easy to analytically calculate $\alpha$ and $\beta$ for a general physical Hamiltonian and noise models, we will numerically show the optimal number of Trotter steps in Sec.~\ref{Sec:numerics}.
In the following, we will first show how to suppress the physical errors $\varepsilon_{phys}$ with the recently proposed physical error mitigation methods. Then, we extend the method to also suppress algorithmic error $\varepsilon_{alg}$.

\section{Error mitigation and extrapolation technique}\label{Sec:errormitigation}
Now, we show how to suppress physical errors with the recently proposed error mitigation methods \cite{Li2017,PhysRevLett.119.180509, endo2017practical, recoveringnoisefree, samerrormitigation,bonet2018low}. Especially, we focus on the linear and exponential extrapolation error mitigation methods \cite{Li2017,PhysRevLett.119.180509,endo2017practical}.
Due to the imperfections of the gate operations, such as decoherence, errors can accumulate in the quantum circuit, so that the noisy output state $\rho_{\mathrm{noise}}$ becomes
\begin{align}
\rho_{\mathrm{noise}}= \mathcal{N}_{N_\mathrm{tot}} \circ  \mathcal{G}_{N_\mathrm{tot}}
\circ\dots  \circ\mathcal{N}_1 \circ  \mathcal{G}_1 (\rho_{\mathrm{init}}),
\end{align}
 where the ideal operation can be expressed as $\mathcal{G}_N \circ \mathcal{G}_{N-1}... \mathcal{G}_1 (\rho_{\mathrm{init}})$. Here, $\mathcal{N}_k$ is the noise channel accompanying the $k^{th}$ ideal gate operation $\mathcal{G}_k$, $N_{\mathrm{tot}}$ is the number of gates, and $\rho_{\mathrm{init}}$ is the initial input state for the quantum circuit.
The noisy output state can be corrected with fault tolerant error correction that
utilises extra qubits to detect errors and correct the state.
However, fault tolerant error correction is considerably costly and is hard to realise with current quantum hardwares. For the noisy intermediate-scale quantum (NISQ) devices with restricted number of qubits, error mitigation methods require no extra qubit and can suppress errors with simple post-processing of different runs of the quantum circuits.

Here we focus on the quantum computation tasks of which the final goal is to calculate the expectation value of a certain observable $\hat{A}$.
Suppose the error-free circuit output state $\rho$, the noiseless expectation value is $\braket{\hat{A}}=\mathrm{Tr}(\hat{A}\rho)$.
For a noisy circuit with error strength $\varepsilon$, the noisy output state ${\rho}_{\varepsilon}$ deviates from the ideal noiseless state. For example, for stochastic errors
\begin{align}
\mathcal{N}_k= (1- p) \mathcal{I} +p \mathcal{N}^\prime_k,
\label{noisee}
\end{align}
we can set $\varepsilon \propto p$, where $\mathcal{I}$ is the identity map, $\mathcal{N}^\prime_k$ is a noise map, and $p\in[0,1]$ denotes the strength of the noise.
The noisy expectation value $\braket{\hat{A}}(\varepsilon)$ can be regarded as a function of $\varepsilon$ and it can be expanded according to different orders of $\varepsilon$,
\begin{align}
\braket{\hat{A}}(\varepsilon)= \braket{\hat{A}}(0)+\sum_{j=1}^n A_j  \varepsilon^j + O(\varepsilon^{n+1}),
\end{align}
{where $A_j$ is  the $j^{th}$ derivative of $\braket{\hat{A}}(\varepsilon)$ in Taylor expansion and $\braket{\hat{A}}(0)$ is the noiseless expectation value. }
For a considerably small $\varepsilon$, we have $\braket{\hat{A}}(\varepsilon) \approx \braket{\hat{A}}(0)$. With several different noisy expectations $\braket{\hat{A}}(\varepsilon)$, error mitigation is to infer the noiseless expectation value $\braket{\hat{A}}(0)$ within a higher accuracy.

\subsection{Linear extrapolation}
Suppose the expectation value of the observable $\braket{\hat{A}}(\varepsilon)$ is measured for several rescaled noise rates $a_j \varepsilon $ with $a_0=1<a_1<a_2...<a_{n}$. Then we can estimate $\braket{\hat{A}}(0)$ by the method introduced in Ref.~\cite{Li2017,PhysRevLett.119.180509},
\begin{equation}
	\begin{aligned}
\braket{\hat{A}}_{est}(0)&= \sum_{i=0}^n \gamma_i \braket{\hat{A}}(a_i \varepsilon),\\
&= \sum_{i=0}^n \gamma_i  \braket{\hat{A}}(0) + \sum_{j=1}^n A_j \varepsilon^j \sum_{i=0}^n \gamma_i a_i ^j + O(\varepsilon^{n+1})\\
&= \braket{\hat{A}}(0) +O(\varepsilon^{n+1}).
\end{aligned}
\end{equation}
Here we choose $\gamma_i$ such that $\sum_{i=0}^n \gamma_i=1 $, $\sum_{i=0}^n \gamma_i a_i ^j =0$ for $j=0, 1, \dots, n$.
By extrapolating $n$ points with different error strengths, we can accurately estimate $\braket{\hat{A}}(0)$ and suppress the error to $O(\varepsilon^{n+1})$. The variance of the estimation $\braket{\hat{A}}_{est}(0)$ is
\begin{align}
\textrm{Var}(\braket{\hat{A}}_{est}(0))= \sum_{i=0}^n \gamma_i ^2\textrm{Var}(\braket{\hat{A}}_{est}(a_i\varepsilon))^2.
\end{align}
Suppose the variance $\textrm{Var}(\braket{\hat{A}}_{est}(a_i\varepsilon))^2$ are similar for different error strengths, the variance of the estimation $\braket{\hat{A}}_{est}(0)$ is $\sum_{i=1}^n \gamma_i^2$ times larger than the variance of each measurement $\hat{A}(\varepsilon)$. Therefore, to achieve the same shot noise of each measurement $\hat{A}(\varepsilon)$, we need to run the circuit $\sum_{i=1}^n \gamma_i^2$ more times and we denote
\begin{align}
\Gamma_{phys}=\sum_{i=1}^n \gamma_i^2
\end{align}
as the cost of physical error mitigation. Note that, $\Gamma_{phys}$ generally increases exponentially to $n$ \cite{PhysRevA.94.062305}. We can thus only choose a small constant of $n$ in practice in order to avoid such exponentially increasing cost.

\subsection{Exponential extrapolation}
The extrapolation method in the previous section is equivalent to using a polynomial function to fit the values with different error strengths and extrapolating the fitting function to estimate the value without error. The optimal fitting function is not necessarily a polynomial function. In Ref.~\cite{endo2017practical},  exponential extrapolation was introduced by fitting an exponential function and was shown to be able to suppress more errors than linear extrapolation. When two error rates, $\epsilon$ and $r \epsilon$ ($r>1$) are used, the estimation value via exponential extrapolation is
\begin{align}
\braket{\hat{A}}_{est}(0)=\braket{\hat{A}}(\epsilon)^{\frac{r}{r-1}} \braket{\hat{A}}(r \epsilon)^{\frac{1}{1-r}}.
\label{exponential}
 \end{align}
 The intuitive reason that the exponential function is a suitable function for extrapolation is as follows. For simplicity, we assume that the noise model for each gate is the same and described by Eq.~(\ref{noisee})
 where $\mathcal{N}^\prime_k $ weakly depends on $p$. Here we can set $\varepsilon=p$ and ignore the gate operations for concise explanation, as it does not have any significance on the argument. Now,  the entire noise process of the quantum circuit can be written as
 \begin{align}
 \prod_{k=1}^{N_{\mathrm{tot}}} \mathcal{N}_k&=\prod_{k=1}^{N_{\mathrm{tot}}} \big[ (1-p) [I] +p \mathcal{N}^\prime _k\big] \\
 &=\sum_{m=0}^{N_{\mathrm{tot}}}  \binom{N_{\mathrm{tot}}}{m} (1-p)^{N_{\mathrm{tot}}-m} p^m \mathcal{K}_m,
 \end{align}
 where $N_{\mathrm{tot}}$ is the number of the gate in the quantum circuit, and $\mathcal{K}_m$ can be expressed as
 \begin{eqnarray}
 \mathcal{K}_m={\binom{N_{\mathrm{tot}}}{m}}^{-1}\times  \left(
\begin{array}{c}
\text{the sum of terms where} \\
\mathcal{N}^\prime \text{ appears for } m \text{ times}
\end{array}
\right).
\end{eqnarray}
The binomial distribution $\binom{N_{\mathrm{tot}}}{m} (1-p)^{N_{\mathrm{tot}}-m} p^m$ can be approximated by a Poisson distribution in the limit of $N_{\mathrm{tot}} \rightarrow \infty $ and $N_{\mathrm{tot}} p \rightarrow \mathrm{const} $, so that
\begin{align}
 \prod_{k=1}^{N_{\mathrm{tot}}} \mathcal{N}_k \approx e^{-N_{\mathrm{tot}} p} \sum_{m=0}^{N_{\mathrm{tot}}} \frac{(N_{\mathrm{tot}}p)^m}{m!} \mathcal{K}_m.
\end{align}
As $\prod_{k=1}^{N_{\mathrm{tot}}} \mathcal{N}_k$ is proportional to $e^{-N_{\mathrm{tot}} p} $, it is suggested in Ref.~\cite{endo2017practical} that the exponential function is a better function for extrapolation.

By assuming that $\braket{\hat{A}}(\epsilon) \propto e^{-N_{\mathrm{tot}} \epsilon} $, and $\textrm{Var}( \braket{\hat{A}} (\epsilon))=\textrm{Var}( \braket{\hat{A}} (r \epsilon))$, we can get the variance of the estimation
\begin{align}
\textrm{Var}(\braket{\hat{A}}_{est}(0)) = \frac{(r^2 e^{2N_{\mathrm{tot}}\epsilon}+e^{2N_{\mathrm{tot}}r\epsilon})}{(r-1)^2} \textrm{Var}( \hat{A} (\epsilon))^2,
\end{align}
where we use the formula for the propagation of the uncertainty. Thus the cost for the exponential extrapolation is approximately
\begin{align}
\Gamma_{phys}=\frac{(r^2 e^{2N_{\mathrm{tot}}\epsilon}+e^{2N_{\mathrm{tot}}r\epsilon})}{(r-1)^2} .
\end{align}

 \section{Error mitigation for algorithmic errors}\label{Sec:algorithmic}
 In this section, we discuss how the algorithmic error in Trotterization can be regarded as an analogy of the physical error and how to extend the extrapolation error mitigation method to suppress the algorithmic error.
Given a Hamiltonian $H$ that has decomposition
\begin{align}
H=\sum_{k} H_k,
\end{align}
the first order Trotter formula approximates the time evolution operator $U(t)=e^{-i\sum_k H_kt}$ by decomposing it into a product form,
\begin{align}\label{trotter}
U(t) = \left( \prod_k e^{-i H_k t/N} \right) ^N + \sum_{i<j}[H_i,H_j]t^2/2N+ \sum_{m=3}^{\infty} E(m),
\end{align}
where $ E(m)$ can be upper bounded by $\|E(m)\|\le N \|Ht/N\|^m/m£¡ $,  and $\|\cdot\|$ denotes the maximal eigenvalue of the matrix. Denote $\varepsilon_N = 1/N$ and $U_{\varepsilon_N} (t)$ as
\begin{equation}
	U_{\varepsilon_N} (t) \equiv \left( \prod_k e^{-i H_k t\varepsilon_N} \right) ^{1/\varepsilon_N},
\end{equation}
then we can straightforwardly see how $U_{\varepsilon_N} (t)$ approximates $U(t)=e^{-i\sum_k H_kt}$, i.e.,
\begin{equation}
	\lim_{\varepsilon_N\rightarrow0^+}U_{\varepsilon_N} (t) =U(t) = e^{-i\sum_k H_kt}.
\end{equation}

Suppose the time evolution operator $U_{\varepsilon_N} (t)$ is applied to an initial state $\ket{\psi_0}$, the output state is $U_{\varepsilon_N} (t)\ket{\psi_0}$. When we measure observable $A$ of the final output state, the average value is
\begin{equation}
	\braket{\hat{A}(t)}(\varepsilon_N)=\bra{\psi_0} U_{\varepsilon_N} (t)^{\dag} \hat{A} U_{\varepsilon_N} (t)\ket{\psi_0}.
\end{equation}
%
While the error-free expectation value for the observable is $\braket{\hat{A} (t)} (0)=\bra{\psi}_0 e^{iHt} \hat{A} e^{-iHt} (t)\ket{\psi}_0$.
Regarding $\braket{\hat{A}(t)}(\varepsilon_N)$ as a function of $\varepsilon_N$,
we can expand $\braket{\hat{A}(t)}(\varepsilon_N)$ as a function of $\varepsilon_N$ by using the Taylor expansion
\begin{equation}
	\braket{\hat{A}(t)}(\varepsilon_N) = \braket{\hat{A} (t)} (0)  +\sum_{j=1}^n \varepsilon_N^j \hat{A} (t)_j + O(\varepsilon_N^{j+1}),
\end{equation}
where $\braket{\hat{A} (t)} (0)$ and $\hat{A} (t)_j$ are independent of $\varepsilon_N$.

Therefore, the extrapolation error mitigation method can be applied to suppress the algorithmic error. In the original error mitigation scheme, we need to boost the error rate $\varepsilon$ to several different error rates $a_j \varepsilon$, which can be realised by intensionally adding more noise to the circuit. Here, as the error rate $\varepsilon_N$ is $1/N$, we can more straightforwardly increase the algorithmic error $\varepsilon_N$ with a smaller number of Trotter steps. For example, by taking $N^\prime =N/2$, we can effectively double the algorithmic error $\varepsilon_{N'} = 2 \varepsilon_N$. With $n$ different numbers of Trotter steps $N_n\le\dots\le N_1\le N_0 = N_{\textrm{opt}}$ and $N_j=N_0/a_j$, we can therefore suppress the algorithmic error by a linear combination of the results
\begin{equation}
	\begin{aligned}
&\braket{\hat{A}(t)}_{est}(0)\\
&= \sum_{i=0}^n \gamma_i^\prime \braket{\hat{A}(t)}(\varepsilon_{N_i})\\
&= \sum_{i=0}^n \gamma_i ^\prime \braket{\hat{A}(t)}(0) + \sum_{j=1}^n \hat{A} (t)_j \varepsilon_{N_{\textrm{opt}}}^j \sum_{i=0}^n \gamma_i ^\prime a_i ^j + O(\varepsilon_{N_{\textrm{opt}}}^{n+1}) \\
&= \braket{\hat{A}(t)}(0) +O(\varepsilon_{N_{\textrm{opt}}}^{n+1}).
\end{aligned}
\end{equation}
Here, $\gamma_i^\prime$ is chosen such that $\sum_{i=0}^n \gamma_i^\prime=1 $, $\sum_{i=0}^n \gamma_i^\prime a_i ^j =0$ for $j=0, 1, \dots, n$. Therefore, we can suppress the algorithmic error to an order of $O((\varepsilon_{N_{\textrm{opt}}})^{n+1})=O((1/{N_{\textrm{opt}}})^{n+1})$. The details of upper bound for algorithmic error are shown in Appendix \ref{appendix: trotter bound}.

The variance of the estimation $\braket{\hat{A}}_{est} (0)$ is
\begin{equation}
\mathrm{Var}\big( \braket{\hat{A}}_{est} (0) \big)= \sum_{i=0}^n \gamma_i^{\prime 2} \mathrm{Var}\big( \braket{\hat{A}}_{est} (\varepsilon_{N_i}) \big)^2.
\end{equation}
Thus, by assuming that the variance is the same for different $\braket{\hat{A}}_{est} (\varepsilon_{N_i})$, the variance of the estimation $\mathrm{Var}\big( \braket{\hat{A}}_{est} (0) \big)$ is $\sum_{i=0}^n \gamma_i^{\prime2}$ times larger than the variance of $\braket{\hat{A}}_{est} (\varepsilon_{N_i})$. We denote the cost of the extrapolation for algorithmic error by
\begin{equation}
	\Gamma_{alg} \equiv \sum_{i=0}^n \gamma_i^{\prime2}.
\end{equation}
Moreover, we can combine the extrapolation for physical errors with the extrapolation for algorithmic errors, and the total cost is
\begin{equation}
	\Gamma_{alg+phys} \equiv \Gamma_{alg}\times\Gamma_{phys}.
\end{equation}

Note that, the only requirement for this method is that  $U_{\varepsilon_N} (t)$ can be expressed as an explicit function of $\varepsilon_N$, and $\lim_{\varepsilon \rightarrow 0^+} U_{\varepsilon_N} (t) =U(t)$.  Therefore, the same argument can be applied to the higher order Trotterisation \cite{suzuki1991general}.
 Furthermore, this method can still be applied even if the expectation value cannot be efficiently expanded as a function of $1/N$. The method also works as long as $\braket{\hat{A}}$ can be expanded with $\varepsilon$ which is a function $f (N_{\mathrm{t}})$ of the tunable parameter $N_{\mathrm{t}}$.

\section{Numerical simulation}\label{Sec:numerics}
\begin{figure}[htb]
\includegraphics[width=3.0cm]{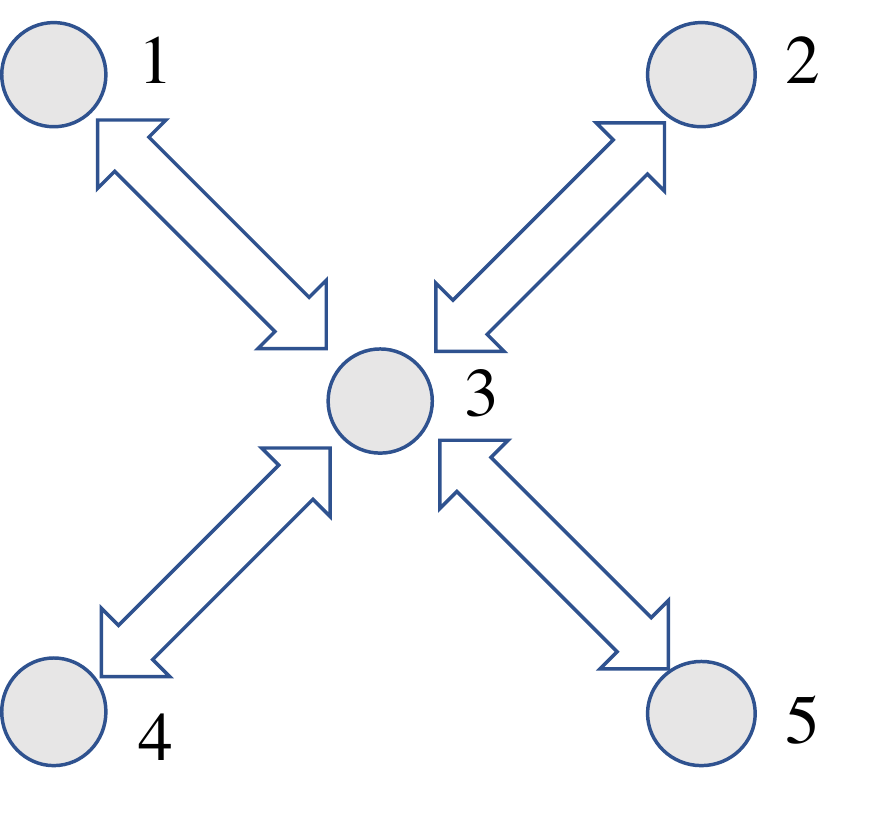}
\caption{Schematic of the five qubit Hamiltonian.}
\label{Fig:hamil}
\end{figure}

In this section, we consider a five qubit system and numerically test our algorithmic error mitigation method for simulating real time evolution.
As shown in shown in Fig.~\ref{Fig:hamil}, the Hamiltonian only has local and near neighborhood interactions,
\begin{align}
H=J \sum_{i=1}  ^5  Z_3 Z_i +B \sum_{i=1}^5 X_i,
\label{hamiltonian}
\end{align}
where $X_i~(Z_i)$ denotes the spin half Pauli $x (z)$ operator acting on the $i$th qubit, $J=3$, and $B=2$. In our numerical simulation, we start with initial state $\ket{\psi_0} = \ket{0,0,0,0,0}$, evolve it with time $t$, and measure $\hat{A} = X_1$.

We consider both algorithmic error from the finite number of Trotter steps and physical error from gate noises.
We consider inhomogeneous Pauli error for both single and two qubit gates,
\begin{equation}\label{inhomogenous}
	\mathcal{E}(\rho)= (1-p) \rho +p_x X \rho X +p_y Y \rho Y +p_z Z \rho Z,
\end{equation}
where $p = p_x+p_y+p_z$.
For single qubit gate, we set $p_x=p_y=2.0 \times 10^{-5}$, $p_z=6.0 \times 10^{-5}$; for two qubit gate, we take error channel $\mathcal{E}_2=\mathcal{E} \otimes \mathcal{E} $ and set  $p_x=p_y=1.0 \times 10^{-4}$, and $p_z=3.0 \times 10^{-4}$ for each $\mathcal{E}$. Note that such a noise rate corresponds to the current state-of-the-art experiment system \cite{PhysRevLett.113.220501,PhysRevLett.117.060504}.

\begin{figure}[htb]
\includegraphics[width=0.95\columnwidth]{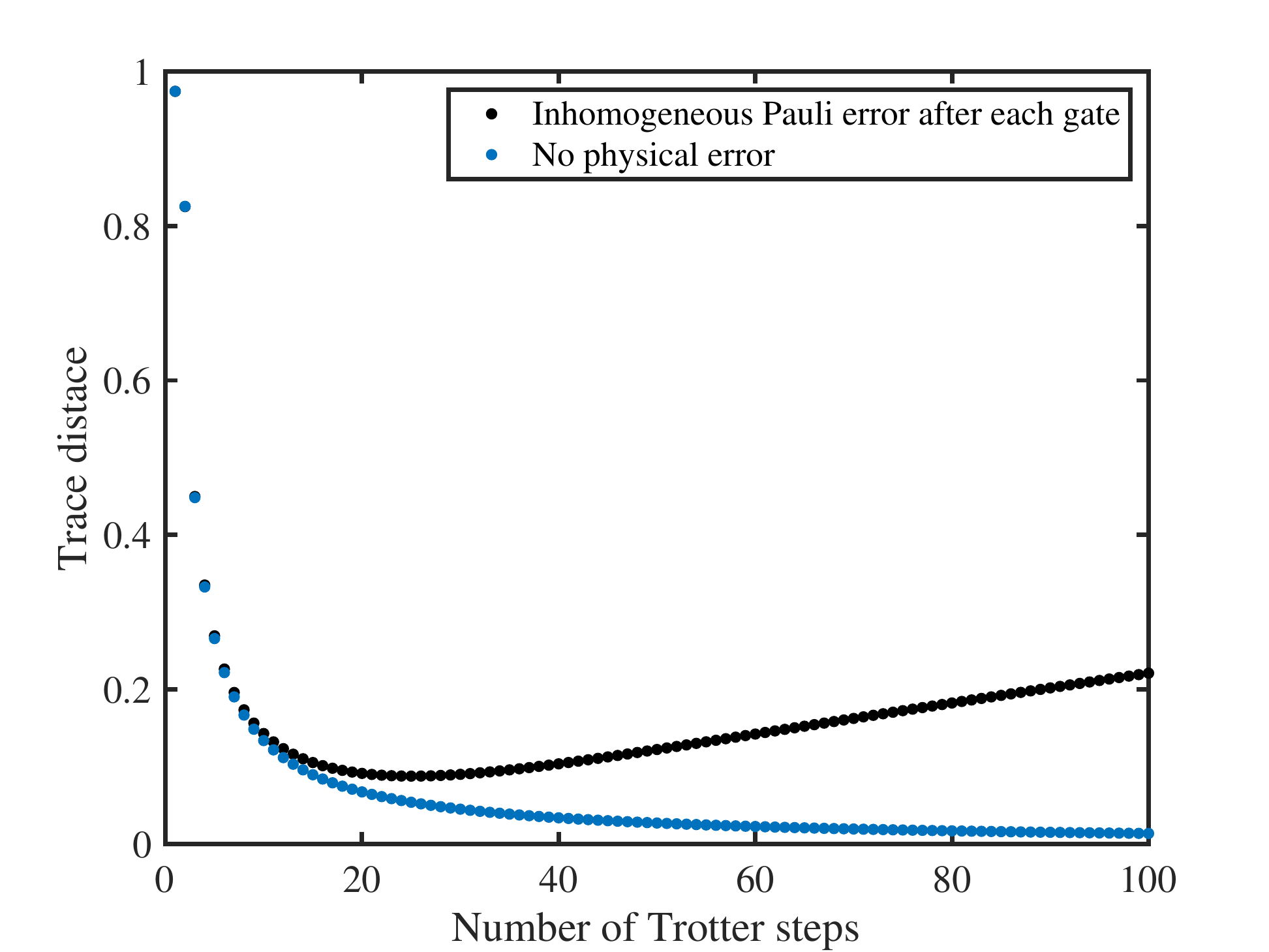}
(a)
\includegraphics[width=0.95\columnwidth]{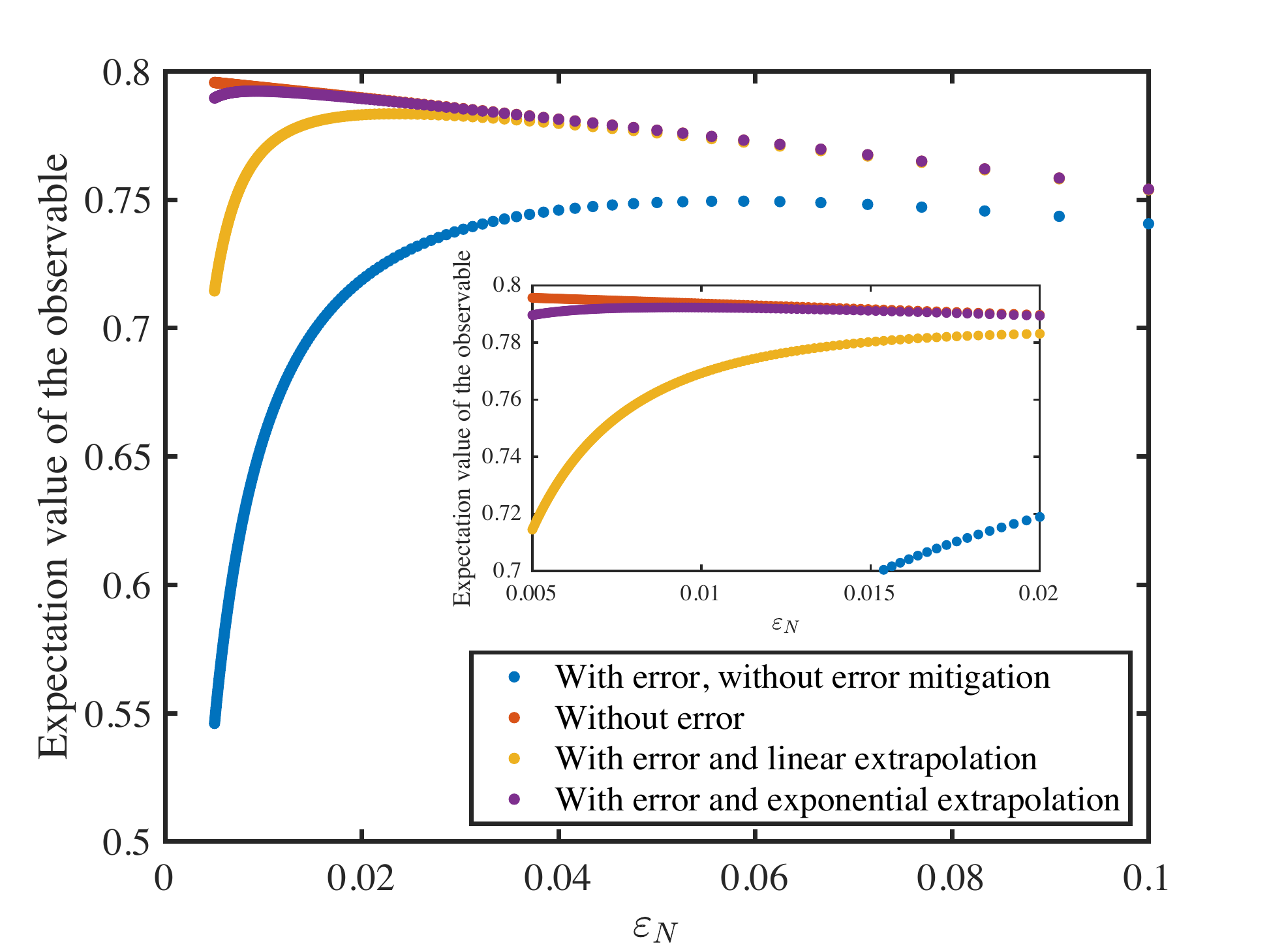}
(b)
\caption{The optimal number of Trotter steps $N$ and the continuity of the measurement as a function of the inverse of the number of Trotter steps $\varepsilon_N = 1/N$.
(a)
The trace distance between the the ideal state and the simulated state in the presence of noise. The blue curve denotes the trace distance without physical error, which goes to zero with an infinite large number of Trotter steps. The black curve denotes the trace distance with inhomogeneous Pauli error after each gate and the optimal number of Trotter steps is $25$.
(b)
The expectation value of the observable versus $\varepsilon_N = 1/N$. Here $N$ is the number of Trotter steps. The horizontal axis corresponds to the number of Trotter step $10-200$. From the simulation result, we can confirm the continuity of the function.} 
\label{optimal}
\end{figure}

Without considering gate error and shot noise to the measurement, one should use the infinite large number of Trotter steps to increase the simulation accuracy. With gate errors, the optimal number of Trotter steps is limited as show in Fig.~\ref{optimal}. Here, we fix the total evolution time $t=0.5$ and numerically find the optimal number of Trotter steps to be $25$.

\begin{figure*}[t]
\minipage{.33\textwidth}
  \includegraphics[width=\linewidth]{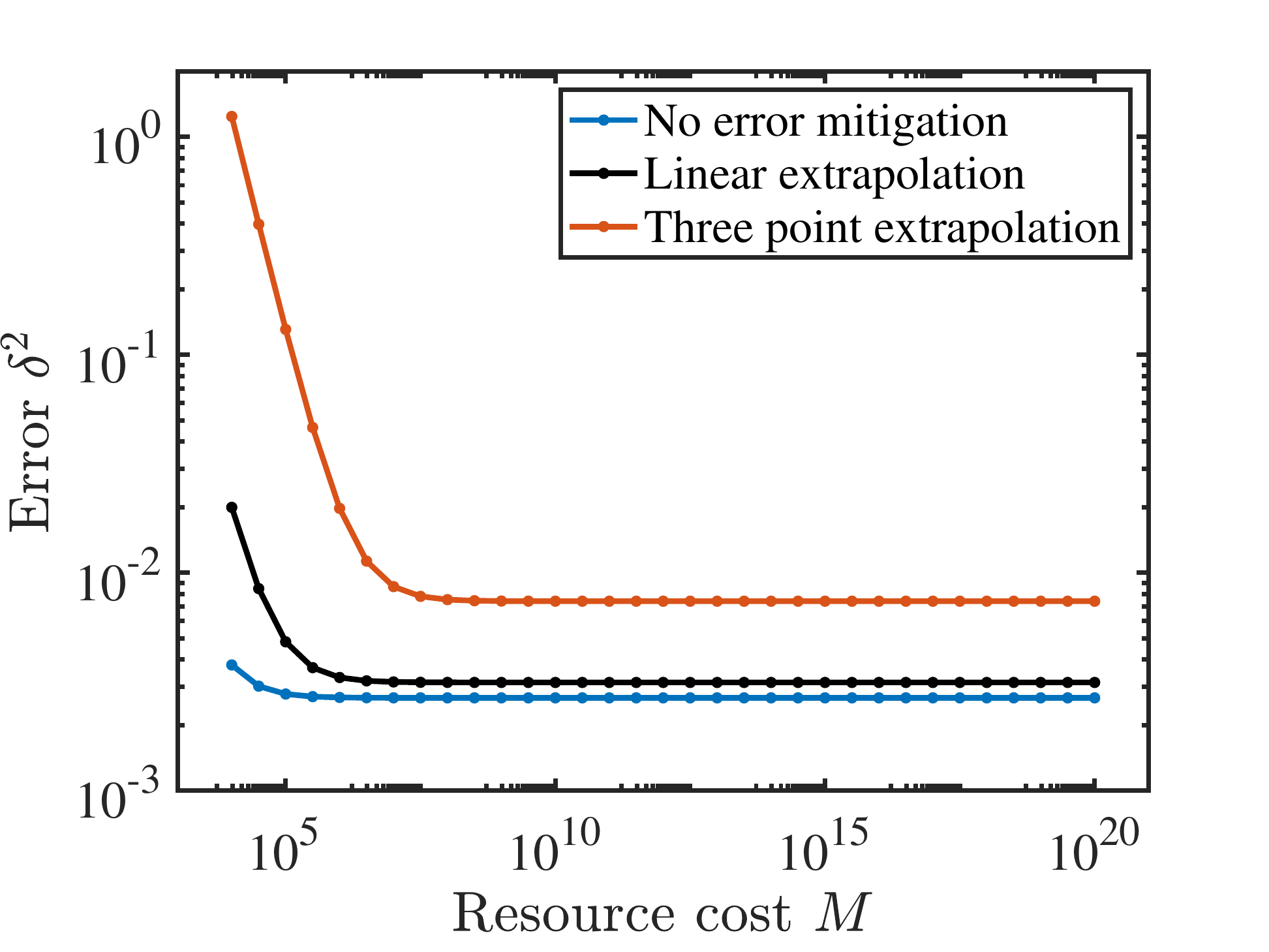}
(a)\label{fig:awesome_image1}
\endminipage\hfill
\minipage{0.33\textwidth}
  \includegraphics[width=\linewidth]{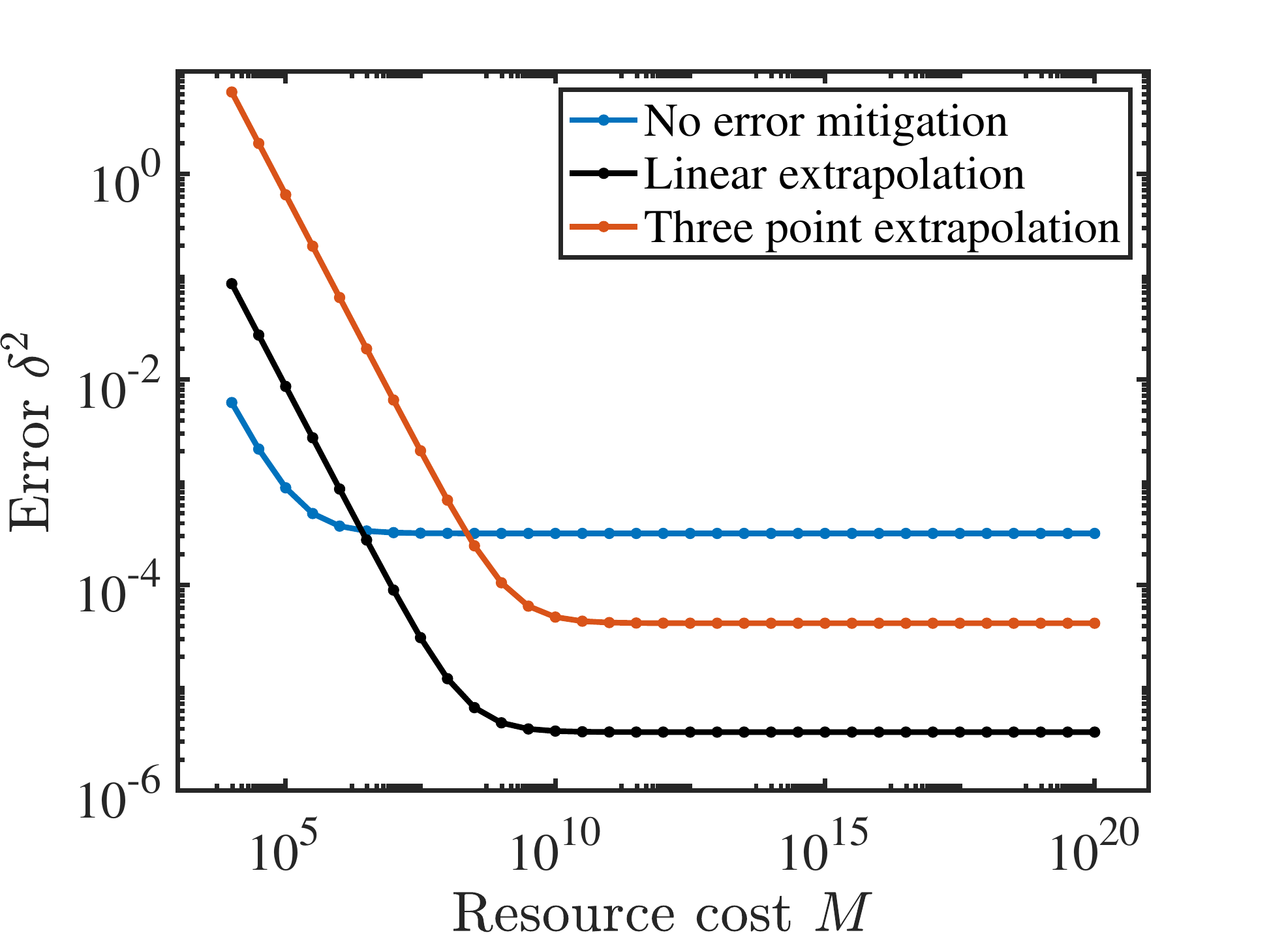}
(b)\label{fig:awesome_image2}
\endminipage\hfill
\minipage{0.33\textwidth}%
  \includegraphics[width=\linewidth]{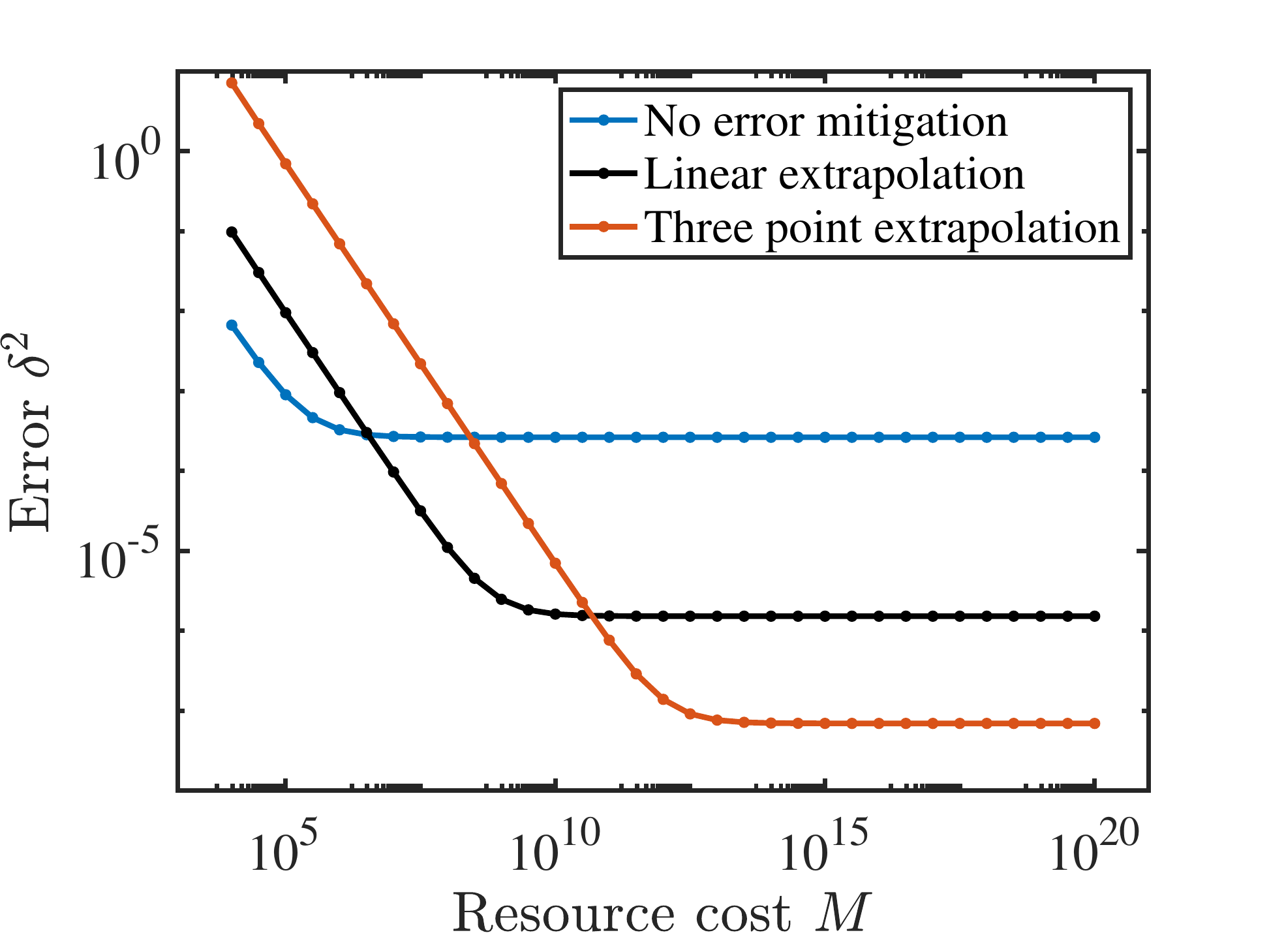}
(c)\label{fig:awesome_image3}
\endminipage
\caption{
The error $\delta^{2}$ against the total runtime resource $M$ for different cases with no algorithmic error mitigation, linear extrapolation for algorithmic errors and three point extrapolation for algorithmic errors. We consider physical errors to the circuit and set the total evolution time  $t=0.5$. The number of Trotter steps $N=25, 20, 15$, $N=25, 15$, $N=25$ are used for the three point, linear extrapolation and no error mitigation case, respectively. Blue line denotes no error mitigation case, black line denotes linear extrapolation for algorithmic errors, red line denotes three point extrapolation for algorithmic error.
(a) No physical error mitigation method is applied. Algorithmic error mitigation cannot improve the simulation accuracy. (b) Linear extrapolation is applied to suppress physical errors. (c) Exponential extrapolation is applied to suppress physical errors.
}
\label{trottermit}
\end{figure*}

To suppress the algorithmic error, we use two or three different number of Trotter steps to infer the value corresponding to the infinite number of Trotter steps.
For a given number of Trotter steps $N$, the runtime of the circuit is related to the circuit depth, which is proportional to $N$, and the circuit repetition time $m$,  which is used to get an accurate estimation of the measurement. We denote the total runtime resource cost $M$ as $M=m N$. To compare different simulation scenarios, we thus consider the same total cost $M$ for a fair comparison.
For three- and two- point extrapolation, we divide $M$ equally to different number of Trotter steps. For example, for three Trotter step extrapolation, $N_1^{(3)}$, $N_2^{(3)}$, and $N_3^{(3)}$, we have $m_i^{(3)} = M/(3N_i^{(3)})$ for $i=1,2,3$; for two Trotter step extrapolation, $N_1^{(2)}$, $N_2^{(2)}$, $m_i^{(2)} = M/(2N_i^{(2)})$ for $i=1,2$; and for one number of Trotter steps $N_1^{(1)}$, $m_1^{(1)} = M/N_1^{(1)}$.

In our simulation, we compare the three cases for suppressing algorithmic errors: no error mitigation, linear extrapolation, and three points extrapolation. In order to quantify the performance of our simulation method, we evaluate the error of the estimation value $\braket{\hat{A}(t)}(0)_{est}$ by
\begin{equation}
\delta^{2} =(\braket{\hat{A}(t)}(0) - \braket{\hat{A}(t)}(0)_{est})^2,
\end{equation}
where $\braket{\hat{A}(t)}(0)$ is the error-free average value.

Now, we show that physical and algorithmic errors can be suppressed with the error mitigation methods.
To begin with, we check the continuity of the $\braket{\hat{A}(t)}$ as a function of $\varepsilon_N = 1/N$, to show that $\braket{\hat{A}(t)}$ can be Taylor expanded by $\varepsilon_N$. We consider the case with and without physical error mitigation. By boosting the error rate twice, and applying the linear and exponential extrapolation methods, we can suppress the physical errors. As shown in Fig.~\ref{optimal} (b), we can see that $\braket{\hat{A}(t)}$ is indeed a continuous function of $\varepsilon_N$, which confirms the possibility of algorithmic error mitigation.
Furthermore, the best accuracy achieved by linear and exponential extrapolation is $\delta^2=2 \times 10^{-4}$ and $2.79 \times 10^{-5}$, with the number of Trotter steps $N=43$ and $109$, respectively.  This clearly shows that we cannot increase the number of Trotter steps infinitely even when the extrapolation method is applied.

To show the effect of algorithmic error mitigation, we also consider three cases for physical errors: no error mitigation, linear extrapolation, and exponential extrapolation.
First, we consider the case where no error mitigation is employed for the physical error. In such a case, physical errors still dominate and we find that algorithmic error extrapolation cannot improve the simulation accuracy, as shown in Fig.~\ref{trottermit}(a).
This is because, although we suppress the algorithmic error by linear combining the results from different number of Trotter steps,  the large deviation due to physical errors even make the estimation worse. {
As the total resource $M$ increases, the shot noise is suppressed and the accuracy converges.}
 The accuracy ${\delta^{2} }$ at the converged point without algorithmic and physical errors mitigations is $2.67\times10^{-3}$.

Next, we consider the cases where linear extrapolation is applied to suppress physical errors.
Subsequently, we apply the algorithmic error extrapolation to suppress algorithmic errors due to Trotterization.  As shown in Fig.~\ref{trottermit}(b),
we find that the algorithmic linear extrapolation outperforms the no error mitigation and three point extrapolation cases for large $M$. The converged accuracy with sufficiently large $M \gtrsim 10^8$ is $\delta^{2}=3.72\times10^{-6}$ under linear extrapolation of algorithmic errors. The improvement of the accuracy is $717$ times,
compared with the case where no error mitigation is applied for both physical and algorithmic errors.

Finally, in Fig.~\ref{trottermit}(c), we plot the result for the case where exponential extrapolation is used for suppressing physical errors. It can be seen that the physical error is successfully reduced, and three point extrapolation works properly, surpassing the performance of linear extrapolation for large $M\gtrsim 10^{13}$. The converged accuracy ${\delta^{2}}$ for sufficiently large $M$ is $6.95 \times 10^{-8}$.
This accuracy correspond to $3.8 \times 10^4$ times improvement, compared with the case where no error mitigation is applied for both physical and algorithmic errors.

\section{Discussion}\label{Sec:discussion}
In this work, we propose an error mitigation method for suppressing algorithmic errors in Trotterization of Hamiltonian simulation. We first show that the optimal number of Trotter steps is finite due to physical errors. Then, we show how the recently proposed physical error mitigation methods can be extended to suppress algorithmic errors in Trotterization. We numerically test our algorithmic error mitigation method in a five qubit Hamiltonian and show how it can improve simulation accuracy by combining it with physical error mitigation.
Although we only focus on the first-order Trotterization, it is worth noting that our scheme can also be extended to other Trotterization schemes, such as higher-order Trotterization \cite{suzuki1991general} and randomisation Trotterization \cite{childs2018faster}. This is because the expectation values obtained from these methods can also be written as a function of the number of Trotter steps. Although, in the presence of physical errors, higher-order Trotter decompositions may not reduce the overall error \cite{Li2017}, we leave these extensions in future works.
Moreover, the other recently proposed Hamiltonian simulation methods, e.g. Taylor series \cite{TaylorSeries} and quantum signal processing \cite{low2016hamiltonian}, offer alternative ways for Hamiltonian simulation instead of Trotterization. These methods  divide the simulation time $t$ into $N$ segments and  simulate the evolution for each time $t/N$. Considering $1/N$ as an error,  our extrapolation method can also be applied by modifying $N$ and extrapolate the results with different $N$ to improve the estimation accuracy.

The extrapolation method has broad applications in diverse fields, including error mitigation, computational chemistry, linear optics simulation, etc.  In classical computational chemistry, extrapolation is widely used in solving molecular structure problem and Monto Carlo simulation of dynamics \cite{brualla2004higher,suzuki1985general}.
In linear optics simulation, the extrapolation method is used to simulate single photons with imperfect photon sources such as coherent states from lasers \cite{PhysRevA.94.062305}.
We can also apply this formalism for mitigating the algorithmic error to other schemes, such as variational quantum eigensolver (VQE) \cite{peruzzo2014variational,PhysRevX.6.031007,kandala2017hardware,VQETheoryNJP,PhysRevA.92.042303,PhysRevA.95.042308,PhysRevX.8.011021}. The result obtained from VQE can be regarded as a function of the depth of the quantum circuit, and  the extrapolation method can be applied.  Similar argument can also be applied to other types of variational algorithms, for simulating real or imaginary time evolutions \cite{Li2017,mcardle2018variational,endo2018discovering}. 

\section*{Acknowledgements.}
We thank SM for insightful discussions. SE is supported by Japan Student
Services Organization (JASSO) Student Exchange Support
Program (Graduate Scholarship for Degree Seeking
Students). This work was supported by BP plc and the EPSRC National Quantum Technology Hub in Networked Quantum Information Technology (EP/M013243/1).

\appendix
\section{Error bound for linear extrapolation}\label{appendix: trotter bound}
Suppose a Hamiltonian $H$ can be decomposed into the sum of local Hamiltonian $H_k$ as $H=\sum_{k} H_k$.
Then, by using the first order Trotter formula, we can approximate the time evolution unitary operators as
\begin{align}
U(t)&=\mathrm{exp} (- i H t) = \bigg( \prod _i ~\mathrm{exp}(-i H_i t/N) \bigg) ^N \\
&+ \sum_{i<j}[H_i,H_j]t^2/2N+ \sum_{m=3}^{\infty} E(m),
\end{align}
where $ E(m)$ can be upper bounded by $\|E(m)\|\le N \|Ht/N\|^m/m! $, $\|\cdot\|$ denotes the maximal eigenvalue. We denote $U_N$ as $\big( \prod _i ~\mathrm{exp}(-i H_i t/N) \big) ^N$, $\varepsilon_N = 1/N$ and $U_{\varepsilon_N} (t)\equiv \left( \prod_k e^{-i H_k t\varepsilon_N} \right) ^{1/\varepsilon_N}$ which converges to $U(t)$. The difference between $U(t)$ and $U_{\varepsilon_N} (t)$ can be rewritten as
  \begin{equation}
\begin{aligned}
U(t)=U_{\varepsilon_N} (t)+a/N+ E^N,
  \end{aligned}
\end{equation}
where $a=\sum_{i<j}[H_i,H_j]t^2/2$  and $E^N=\sum_{m=3}^{\infty} E(m)$ which can be bound by
\begin{equation}
\begin{aligned}
\|E^N\|&\le N\sum_{m=3}^{\infty} \|Ht/N\|^m/m!\\
&<\frac{\|H\|^3t^3}{6N^2} e^{\|H\|t/N}.
\end{aligned}
\end{equation}

Suppose the time evolution operator $U_{\varepsilon_N} (t)$ is applied to an initial state $\ket{\psi_0}$, the output state is $U_{\varepsilon_N} (t)\ket{\psi_0}$. When we measure observable $A$ of the final output state, the average value is
\begin{equation}
	\braket{\hat{A}(t)}(\varepsilon_N)=\bra{\psi_0} U_{\varepsilon_N} (t)^{\dag} \hat{A} U_{\varepsilon_N} (t)\ket{\psi_0}.
\end{equation}
and the average value for evolution $U(t)$ is $\bra{\psi_0} e^{iHt} \hat{A} e^{-iHt} \ket{\psi_0}$. The relationship between them can be expressed as
\begin{equation}
\begin{aligned}
&\braket{\hat{A}(t)}(\varepsilon_N)\\
=&\braket{\hat{A}(t)}(0)- 1/N[\bra{\psi_0} U^{\dag} \hat{A} a\ket{\psi_0}+\bra{\psi_0} a^{\dag} \hat{A} U\ket{\psi_0}]\\
&+\bra{\psi_0} (a/N+E^N)^{\dag} \hat{A} (a/N+E^N)\ket{\psi_0}\\
&-[\bra{\Psi} U^{\dag} \hat{A} E^N\ket{\Psi}+\bra{\Psi}(E^N)^{\dag} \hat{A} U\ket{\Psi}]\\
=&\braket{\hat{A}(t)}(0)+ b/N+ \mathcal{R}^N
\end{aligned}
\end{equation}
where $b=-[\bra{\psi_0} U^{\dag} \hat{A} a\ket{\psi_0}+\bra{\psi_0} a^{\dag} \hat{A} U\ket{\psi_0}]$, independent of $N$, and $\mathcal{R}^N=\bra{\psi_0} (a/N+E^N)^{\dag} \hat{A} (a/N+E^N)\ket{\psi_0}- [\bra{\psi_0} U^{\dag} \hat{A} E^N\ket{\psi_0}+\bra{\psi_0}(E^N)^{\dag} \hat{A} U\ket{\psi_0}]$ is bound by
 \begin{equation}
\begin{aligned}
\mathcal{R}^N \le [(\frac{\|a\|}{N}+\frac{\|H\|^3t^3}{6N^2} e^{\|H\|t/N})^2+\frac{\|H\|^3t^3}{3N^2} e^{\|H\|t/N}] \|\hat{A}\|.
 \end{aligned}
 \end{equation}

 For simplicity, we rewrite the upper bound as
 \begin{equation}
\begin{aligned}
 \mathcal{R}_2^N\le \frac{\|a\|^2+\|H\|^3t^3 e^{\|H\|t/N}/3 }{N^2}\|\hat{A}\|+\lambda_N,
 \end{aligned}
\end{equation}
where $\lambda_N$ is the summation of other high order terms and the bound for $\mathcal{R}^N$ is  $O(\frac{1}{N^2})$.

Now we use the linear combination of $\braket{\hat{A}(t)}(\varepsilon_{N_1})$ and $\braket{\hat{A}(t)}(\varepsilon_{N_2})$ to estimate $\braket{\hat{A}(t)}(0)$ by

\begin{equation}
\begin{aligned}
\braket{\hat{A}(t)}_{est}(0)= \beta_1 \braket{\hat{A}(t)}(\varepsilon_{N_1})+\beta_2 \braket{\hat{A}(t)}(\varepsilon_{N_2}).
\end{aligned}
\end{equation}
We choose the parameter $\beta_1,\beta_2$ satisfying $\beta_1 /N_1+\beta_2/N_2=0$ and $\beta_1+\beta_2 =1$  so that the difference between $\braket{\hat{A}(t)}_{est}(0)$ and $\braket{\hat{A}(t)}(0)$ is

\begin{equation}
\begin{aligned}
&|\braket{\hat{A}(t)}_{est}(0)-\braket{\hat{A}(t)}(0)|\\
&\le  \mathcal{R}^{N_1}\beta_1+   \mathcal{R}^{N_2}\beta_2\\
  &\le  \sum_{n=N_1, N_2} (\frac{\|a\|^2+\|H\|^3t^3 e^{\|H\|t/n}/3 }{n^2}\|\hat{A}\|+\lambda_n )
\end{aligned}
\end{equation}
Based on the two point extrapolation technique, we use the simulation results with Trotter steps $N_1$ $N_2$, which both are $O(N)$ to
reduce the error from  $O(\frac{1}{N})$ to $O(\frac{1}{N^2})$.
Similar to the above analysis, by applying extrapolation with $m$ different points, we can suppress the Trotter error to the order of $O(1/N^m)$.

\bibliographystyle{apsrev4-1}
\bibliography{bibTrotter}

\begin{thebibliography}{32}%
\makeatletter
\providecommand \@ifxundefined [1]{%
 \@ifx{#1\undefined}
}%
\providecommand \@ifnum [1]{%
 \ifnum #1\expandafter \@firstoftwo
 \else \expandafter \@secondoftwo
 \fi
}%
\providecommand \@ifx [1]{%
 \ifx #1\expandafter \@firstoftwo
 \else \expandafter \@secondoftwo
 \fi
}%
\providecommand \natexlab [1]{#1}%
\providecommand \enquote  [1]{``#1''}%
\providecommand \bibnamefont  [1]{#1}%
\providecommand \bibfnamefont [1]{#1}%
\providecommand \citenamefont [1]{#1}%
\providecommand \href@noop [0]{\@secondoftwo}%
\providecommand \href [0]{\begingroup \@sanitize@url \@href}%
\providecommand \@href[1]{\@@startlink{#1}\@@href}%
\providecommand \@@href[1]{\endgroup#1\@@endlink}%
\providecommand \@sanitize@url [0]{\catcode `\\12\catcode `\$12\catcode
  `\&12\catcode `\#12\catcode `\^12\catcode `\_12\catcode `\%12\relax}%
\providecommand \@@startlink[1]{}%
\providecommand \@@endlink[0]{}%
\providecommand \url  [0]{\begingroup\@sanitize@url \@url }%
\providecommand \@url [1]{\endgroup\@href {#1}{\urlprefix }}%
\providecommand \urlprefix  [0]{URL }%
\providecommand \Eprint [0]{\href }%
\providecommand \doibase [0]{http://dx.doi.org/}%
\providecommand \selectlanguage [0]{\@gobble}%
\providecommand \bibinfo  [0]{\@secondoftwo}%
\providecommand \bibfield  [0]{\@secondoftwo}%
\providecommand \translation [1]{[#1]}%
\providecommand \BibitemOpen [0]{}%
\providecommand \bibitemStop [0]{}%
\providecommand \bibitemNoStop [0]{.\EOS\space}%
\providecommand \EOS [0]{\spacefactor3000\relax}%
\providecommand \BibitemShut  [1]{\csname bibitem#1\endcsname}%
\let\auto@bib@innerbib\@empty
\bibitem [{\citenamefont {Feynman}(1982)}]{Feynman1982}%
  \BibitemOpen
  \bibfield  {author} {\bibinfo {author} {\bibfnamefont {R.~P.}\ \bibnamefont
  {Feynman}},\ }\href {\doibase 10.1007/BF02650179} {\bibfield  {journal}
  {\bibinfo  {journal} {International Journal of Theoretical Physics}\ }\textbf
  {\bibinfo {volume} {21}},\ \bibinfo {pages} {467} (\bibinfo {year}
  {1982})}\BibitemShut {NoStop}%
\bibitem [{\citenamefont {Lloyd}(1996)}]{sethuniversal}%
  \BibitemOpen
  \bibfield  {author} {\bibinfo {author} {\bibfnamefont {S.}~\bibnamefont
  {Lloyd}},\ }\href {\doibase https://doi.org/10.1126/science.273.5278.1073}
  {\bibfield  {journal} {\bibinfo  {journal} {Science}\ }\textbf {\bibinfo
  {volume} {273}},\ \bibinfo {pages} {1073} (\bibinfo {year}
  {1996})}\BibitemShut {NoStop}%
\bibitem [{\citenamefont {Aspuru-Guzik}\ \emph {et~al.}(2005)\citenamefont
  {Aspuru-Guzik}, \citenamefont {Dutoi}, \citenamefont {Love},\ and\
  \citenamefont {Head-Gordon}}]{aspuru2005simulated}%
  \BibitemOpen
  \bibfield  {author} {\bibinfo {author} {\bibfnamefont {A.}~\bibnamefont
  {Aspuru-Guzik}}, \bibinfo {author} {\bibfnamefont {A.~D.}\ \bibnamefont
  {Dutoi}}, \bibinfo {author} {\bibfnamefont {P.~J.}\ \bibnamefont {Love}}, \
  and\ \bibinfo {author} {\bibfnamefont {M.}~\bibnamefont {Head-Gordon}},\
  }\href@noop {} {\bibfield  {journal} {\bibinfo  {journal} {Science}\ }\textbf
  {\bibinfo {volume} {309}},\ \bibinfo {pages} {1704} (\bibinfo {year}
  {2005})}\BibitemShut {NoStop}%
\bibitem [{\citenamefont {Suzuki}(1991)}]{suzuki1991general}%
  \BibitemOpen
  \bibfield  {author} {\bibinfo {author} {\bibfnamefont {M.}~\bibnamefont
  {Suzuki}},\ }\href@noop {} {\bibfield  {journal} {\bibinfo  {journal}
  {Journal of Mathematical Physics}\ }\textbf {\bibinfo {volume} {32}},\
  \bibinfo {pages} {400} (\bibinfo {year} {1991})}\BibitemShut {NoStop}%
\bibitem [{\citenamefont {Childs}\ \emph {et~al.}(2018)\citenamefont {Childs},
  \citenamefont {Ostrander},\ and\ \citenamefont {Su}}]{childs2018faster}%
  \BibitemOpen
  \bibfield  {author} {\bibinfo {author} {\bibfnamefont {A.~M.}\ \bibnamefont
  {Childs}}, \bibinfo {author} {\bibfnamefont {A.}~\bibnamefont {Ostrander}}, \
  and\ \bibinfo {author} {\bibfnamefont {Y.}~\bibnamefont {Su}},\ }\href@noop
  {} {\bibfield  {journal} {\bibinfo  {journal} {arXiv preprint
  arXiv:1805.08385}\ } (\bibinfo {year} {2018})}\BibitemShut {NoStop}%
\bibitem [{\citenamefont {Berry}\ \emph {et~al.}(2015)\citenamefont {Berry},
  \citenamefont {Childs}, \citenamefont {Cleve}, \citenamefont {Kothari},\ and\
  \citenamefont {Somma}}]{TaylorSeries}%
  \BibitemOpen
  \bibfield  {author} {\bibinfo {author} {\bibfnamefont {D.~W.}\ \bibnamefont
  {Berry}}, \bibinfo {author} {\bibfnamefont {A.~M.}\ \bibnamefont {Childs}},
  \bibinfo {author} {\bibfnamefont {R.}~\bibnamefont {Cleve}}, \bibinfo
  {author} {\bibfnamefont {R.}~\bibnamefont {Kothari}}, \ and\ \bibinfo
  {author} {\bibfnamefont {R.~D.}\ \bibnamefont {Somma}},\ }\href {\doibase
  10.1103/PhysRevLett.114.090502} {\bibfield  {journal} {\bibinfo  {journal}
  {Phys. Rev. Lett.}\ }\textbf {\bibinfo {volume} {114}},\ \bibinfo {pages}
  {090502} (\bibinfo {year} {2015})}\BibitemShut {NoStop}%
\bibitem [{\citenamefont {Low}\ and\ \citenamefont
  {Chuang}(2016)}]{low2016hamiltonian}%
  \BibitemOpen
  \bibfield  {author} {\bibinfo {author} {\bibfnamefont {G.~H.}\ \bibnamefont
  {Low}}\ and\ \bibinfo {author} {\bibfnamefont {I.~L.}\ \bibnamefont
  {Chuang}},\ }\href@noop {} {\bibfield  {journal} {\bibinfo  {journal} {arXiv
  preprint arXiv:1610.06546}\ } (\bibinfo {year} {2016})}\BibitemShut {NoStop}%
\bibitem [{\citenamefont {Childs}\ \emph {et~al.}(2017)\citenamefont {Childs},
  \citenamefont {Maslov}, \citenamefont {Nam}, \citenamefont {Ross},\ and\
  \citenamefont {Su}}]{childs2017toward}%
  \BibitemOpen
  \bibfield  {author} {\bibinfo {author} {\bibfnamefont {A.~M.}\ \bibnamefont
  {Childs}}, \bibinfo {author} {\bibfnamefont {D.}~\bibnamefont {Maslov}},
  \bibinfo {author} {\bibfnamefont {Y.}~\bibnamefont {Nam}}, \bibinfo {author}
  {\bibfnamefont {N.~J.}\ \bibnamefont {Ross}}, \ and\ \bibinfo {author}
  {\bibfnamefont {Y.}~\bibnamefont {Su}},\ }\href@noop {} {\bibfield  {journal}
  {\bibinfo  {journal} {arXiv preprint arXiv:1711.10980}\ } (\bibinfo {year}
  {2017})}\BibitemShut {NoStop}%
\bibitem [{\citenamefont {Suzuki}(1976)}]{trottersuzuki}%
  \BibitemOpen
  \bibfield  {author} {\bibinfo {author} {\bibfnamefont {M.}~\bibnamefont
  {Suzuki}},\ }\href@noop {} {\bibfield  {journal} {\bibinfo  {journal} {Comm.
  Math. Phys.}\ }\textbf {\bibinfo {volume} {51}},\ \bibinfo {pages} {183}
  (\bibinfo {year} {1976})}\BibitemShut {NoStop}%
\bibitem [{\citenamefont {Preskill}(2018)}]{preskill2018quantum}%
  \BibitemOpen
  \bibfield  {author} {\bibinfo {author} {\bibfnamefont {J.}~\bibnamefont
  {Preskill}},\ }\href@noop {} {\bibfield  {journal} {\bibinfo  {journal}
  {arXiv preprint arXiv:1801.00862}\ } (\bibinfo {year} {2018})}\BibitemShut
  {NoStop}%
\bibitem [{\citenamefont {Knee}\ and\ \citenamefont {Munro}(2015)}]{george}%
  \BibitemOpen
  \bibfield  {author} {\bibinfo {author} {\bibfnamefont {G.~C.}\ \bibnamefont
  {Knee}}\ and\ \bibinfo {author} {\bibfnamefont {W.~J.}\ \bibnamefont
  {Munro}},\ }\href {\doibase 10.1103/PhysRevA.91.052327} {\bibfield  {journal}
  {\bibinfo  {journal} {Phys. Rev. A}\ }\textbf {\bibinfo {volume} {91}},\
  \bibinfo {pages} {052327} (\bibinfo {year} {2015})}\BibitemShut {NoStop}%
\bibitem [{\citenamefont {Li}\ and\ \citenamefont {Benjamin}(2017)}]{Li2017}%
  \BibitemOpen
  \bibfield  {author} {\bibinfo {author} {\bibfnamefont {Y.}~\bibnamefont
  {Li}}\ and\ \bibinfo {author} {\bibfnamefont {S.~C.}\ \bibnamefont
  {Benjamin}},\ }\href {\doibase 10.1103/PhysRevX.7.021050} {\bibfield
  {journal} {\bibinfo  {journal} {Phys. Rev. X}\ }\textbf {\bibinfo {volume}
  {7}},\ \bibinfo {pages} {021050} (\bibinfo {year} {2017})}\BibitemShut
  {NoStop}%
\bibitem [{\citenamefont {Temme}\ \emph {et~al.}(2017)\citenamefont {Temme},
  \citenamefont {Bravyi},\ and\ \citenamefont
  {Gambetta}}]{PhysRevLett.119.180509}%
  \BibitemOpen
  \bibfield  {author} {\bibinfo {author} {\bibfnamefont {K.}~\bibnamefont
  {Temme}}, \bibinfo {author} {\bibfnamefont {S.}~\bibnamefont {Bravyi}}, \
  and\ \bibinfo {author} {\bibfnamefont {J.~M.}\ \bibnamefont {Gambetta}},\
  }\href {\doibase 10.1103/PhysRevLett.119.180509} {\bibfield  {journal}
  {\bibinfo  {journal} {Phys. Rev. Lett.}\ }\textbf {\bibinfo {volume} {119}},\
  \bibinfo {pages} {180509} (\bibinfo {year} {2017})}\BibitemShut {NoStop}%
\bibitem [{\citenamefont {Endo}\ \emph
  {et~al.}(2018{\natexlab{a}})\citenamefont {Endo}, \citenamefont {Benjamin},\
  and\ \citenamefont {Li}}]{endo2017practical}%
  \BibitemOpen
  \bibfield  {author} {\bibinfo {author} {\bibfnamefont {S.}~\bibnamefont
  {Endo}}, \bibinfo {author} {\bibfnamefont {S.~C.}\ \bibnamefont {Benjamin}},
  \ and\ \bibinfo {author} {\bibfnamefont {Y.}~\bibnamefont {Li}},\ }\href@noop
  {} {\bibfield  {journal} {\bibinfo  {journal} {Phys. Rev. X}\ }\textbf
  {\bibinfo {volume} {8}},\ \bibinfo {pages} {031027} (\bibinfo {year}
  {2018}{\natexlab{a}})}\BibitemShut {NoStop}%
\bibitem [{\citenamefont {Otten}\ and\ \citenamefont
  {Gray}(2018)}]{recoveringnoisefree}%
  \BibitemOpen
  \bibfield  {author} {\bibinfo {author} {\bibfnamefont {M.}~\bibnamefont
  {Otten}}\ and\ \bibinfo {author} {\bibfnamefont {S.}~\bibnamefont {Gray}},\
  }\href@noop {} {\bibfield  {journal} {\bibinfo  {journal} {arXiv preprint
  arXiv:1806.07860}\ } (\bibinfo {year} {2018})}\BibitemShut {NoStop}%
\bibitem [{\citenamefont {McArdle}\ \emph
  {et~al.}(2018{\natexlab{a}})\citenamefont {McArdle}, \citenamefont {Yuan},\
  and\ \citenamefont {Benjamin}}]{samerrormitigation}%
  \BibitemOpen
  \bibfield  {author} {\bibinfo {author} {\bibfnamefont {S.}~\bibnamefont
  {McArdle}}, \bibinfo {author} {\bibfnamefont {X.}~\bibnamefont {Yuan}}, \
  and\ \bibinfo {author} {\bibfnamefont {S.}~\bibnamefont {Benjamin}},\
  }\href@noop {} {\bibfield  {journal} {\bibinfo  {journal} {arXiv preprint
  arXiv: 1807.02467}\ } (\bibinfo {year} {2018}{\natexlab{a}})}\BibitemShut
  {NoStop}%
\bibitem [{\citenamefont {Bonet-Monroig}\ \emph {et~al.}(2018)\citenamefont
  {Bonet-Monroig}, \citenamefont {Sagastizabal}, \citenamefont {Singh},\ and\
  \citenamefont {O'Brien}}]{bonet2018low}%
  \BibitemOpen
  \bibfield  {author} {\bibinfo {author} {\bibfnamefont {X.}~\bibnamefont
  {Bonet-Monroig}}, \bibinfo {author} {\bibfnamefont {R.}~\bibnamefont
  {Sagastizabal}}, \bibinfo {author} {\bibfnamefont {M.}~\bibnamefont {Singh}},
  \ and\ \bibinfo {author} {\bibfnamefont {T.}~\bibnamefont {O'Brien}},\
  }\href@noop {} {\bibfield  {journal} {\bibinfo  {journal} {arXiv preprint
  arXiv:1807.10050}\ } (\bibinfo {year} {2018})}\BibitemShut {NoStop}%
\bibitem [{\citenamefont {Fuchs}(1996)}]{distinguishability}%
  \BibitemOpen
  \bibfield  {author} {\bibinfo {author} {\bibfnamefont {C.~A.}\ \bibnamefont
  {Fuchs}},\ }\href@noop {} {\bibfield  {journal} {\bibinfo  {journal} {arXiv
  preprint quant-ph/9601020}\ } (\bibinfo {year} {1996})}\BibitemShut {NoStop}%
\bibitem [{\citenamefont {Yuan}\ \emph {et~al.}(2016)\citenamefont {Yuan},
  \citenamefont {Zhang}, \citenamefont {L\"utkenhaus},\ and\ \citenamefont
  {Ma}}]{PhysRevA.94.062305}%
  \BibitemOpen
  \bibfield  {author} {\bibinfo {author} {\bibfnamefont {X.}~\bibnamefont
  {Yuan}}, \bibinfo {author} {\bibfnamefont {Z.}~\bibnamefont {Zhang}},
  \bibinfo {author} {\bibfnamefont {N.}~\bibnamefont {L\"utkenhaus}}, \ and\
  \bibinfo {author} {\bibfnamefont {X.}~\bibnamefont {Ma}},\ }\href {\doibase
  10.1103/PhysRevA.94.062305} {\bibfield  {journal} {\bibinfo  {journal} {Phys.
  Rev. A}\ }\textbf {\bibinfo {volume} {94}},\ \bibinfo {pages} {062305}
  (\bibinfo {year} {2016})}\BibitemShut {NoStop}%
\bibitem [{\citenamefont {Harty}\ \emph {et~al.}(2014)\citenamefont {Harty},
  \citenamefont {Allcock}, \citenamefont {Ballance}, \citenamefont {Guidoni},
  \citenamefont {Janacek}, \citenamefont {Linke}, \citenamefont {Stacey},\ and\
  \citenamefont {Lucas}}]{PhysRevLett.113.220501}%
  \BibitemOpen
  \bibfield  {author} {\bibinfo {author} {\bibfnamefont {T.~P.}\ \bibnamefont
  {Harty}}, \bibinfo {author} {\bibfnamefont {D.~T.~C.}\ \bibnamefont
  {Allcock}}, \bibinfo {author} {\bibfnamefont {C.~J.}\ \bibnamefont
  {Ballance}}, \bibinfo {author} {\bibfnamefont {L.}~\bibnamefont {Guidoni}},
  \bibinfo {author} {\bibfnamefont {H.~A.}\ \bibnamefont {Janacek}}, \bibinfo
  {author} {\bibfnamefont {N.~M.}\ \bibnamefont {Linke}}, \bibinfo {author}
  {\bibfnamefont {D.~N.}\ \bibnamefont {Stacey}}, \ and\ \bibinfo {author}
  {\bibfnamefont {D.~M.}\ \bibnamefont {Lucas}},\ }\href {\doibase
  10.1103/PhysRevLett.113.220501} {\bibfield  {journal} {\bibinfo  {journal}
  {Phys. Rev. Lett.}\ }\textbf {\bibinfo {volume} {113}},\ \bibinfo {pages}
  {220501} (\bibinfo {year} {2014})}\BibitemShut {NoStop}%
\bibitem [{\citenamefont {Ballance}\ \emph {et~al.}(2016)\citenamefont
  {Ballance}, \citenamefont {Harty}, \citenamefont {Linke}, \citenamefont
  {Sepiol},\ and\ \citenamefont {Lucas}}]{PhysRevLett.117.060504}%
  \BibitemOpen
  \bibfield  {author} {\bibinfo {author} {\bibfnamefont {C.~J.}\ \bibnamefont
  {Ballance}}, \bibinfo {author} {\bibfnamefont {T.~P.}\ \bibnamefont {Harty}},
  \bibinfo {author} {\bibfnamefont {N.~M.}\ \bibnamefont {Linke}}, \bibinfo
  {author} {\bibfnamefont {M.~A.}\ \bibnamefont {Sepiol}}, \ and\ \bibinfo
  {author} {\bibfnamefont {D.~M.}\ \bibnamefont {Lucas}},\ }\href {\doibase
  10.1103/PhysRevLett.117.060504} {\bibfield  {journal} {\bibinfo  {journal}
  {Phys. Rev. Lett.}\ }\textbf {\bibinfo {volume} {117}},\ \bibinfo {pages}
  {060504} (\bibinfo {year} {2016})}\BibitemShut {NoStop}%
\bibitem [{\citenamefont {Brualla}\ \emph {et~al.}(2004)\citenamefont
  {Brualla}, \citenamefont {Sakkos}, \citenamefont {Boronat},\ and\
  \citenamefont {Casulleras}}]{brualla2004higher}%
  \BibitemOpen
  \bibfield  {author} {\bibinfo {author} {\bibfnamefont {L.}~\bibnamefont
  {Brualla}}, \bibinfo {author} {\bibfnamefont {K.}~\bibnamefont {Sakkos}},
  \bibinfo {author} {\bibfnamefont {J.}~\bibnamefont {Boronat}}, \ and\
  \bibinfo {author} {\bibfnamefont {J.}~\bibnamefont {Casulleras}},\
  }\href@noop {} {\bibfield  {journal} {\bibinfo  {journal} {The Journal of
  chemical physics}\ }\textbf {\bibinfo {volume} {121}},\ \bibinfo {pages}
  {636} (\bibinfo {year} {2004})}\BibitemShut {NoStop}%
\bibitem [{\citenamefont {Suzuki}(1985)}]{suzuki1985general}%
  \BibitemOpen
  \bibfield  {author} {\bibinfo {author} {\bibfnamefont {M.}~\bibnamefont
  {Suzuki}},\ }\href@noop {} {\bibfield  {journal} {\bibinfo  {journal}
  {Physics Letters A}\ }\textbf {\bibinfo {volume} {113}},\ \bibinfo {pages}
  {299} (\bibinfo {year} {1985})}\BibitemShut {NoStop}%
\bibitem [{\citenamefont {Peruzzo}\ \emph {et~al.}(2014)\citenamefont
  {Peruzzo}, \citenamefont {McClean}, \citenamefont {Shadbolt}, \citenamefont
  {Yung}, \citenamefont {Zhou}, \citenamefont {Love}, \citenamefont
  {Aspuru-Guzik},\ and\ \citenamefont {O’brien}}]{peruzzo2014variational}%
  \BibitemOpen
  \bibfield  {author} {\bibinfo {author} {\bibfnamefont {A.}~\bibnamefont
  {Peruzzo}}, \bibinfo {author} {\bibfnamefont {J.}~\bibnamefont {McClean}},
  \bibinfo {author} {\bibfnamefont {P.}~\bibnamefont {Shadbolt}}, \bibinfo
  {author} {\bibfnamefont {M.-H.}\ \bibnamefont {Yung}}, \bibinfo {author}
  {\bibfnamefont {X.-Q.}\ \bibnamefont {Zhou}}, \bibinfo {author}
  {\bibfnamefont {P.~J.}\ \bibnamefont {Love}}, \bibinfo {author}
  {\bibfnamefont {A.}~\bibnamefont {Aspuru-Guzik}}, \ and\ \bibinfo {author}
  {\bibfnamefont {J.~L.}\ \bibnamefont {O’brien}},\ }\href@noop {} {\bibfield
   {journal} {\bibinfo  {journal} {Nature communications}\ }\textbf {\bibinfo
  {volume} {5}} (\bibinfo {year} {2014})}\BibitemShut {NoStop}%
\bibitem [{\citenamefont {O'Malley}\ \emph {et~al.}(2016)\citenamefont
  {O'Malley}, \citenamefont {Babbush}, \citenamefont {Kivlichan}, \citenamefont
  {Romero}, \citenamefont {McClean}, \citenamefont {Barends}, \citenamefont
  {Kelly}, \citenamefont {Roushan}, \citenamefont {Tranter}, \citenamefont
  {Ding}, \citenamefont {Campbell}, \citenamefont {Chen}, \citenamefont {Chen},
  \citenamefont {Chiaro}, \citenamefont {Dunsworth}, \citenamefont {Fowler},
  \citenamefont {Jeffrey}, \citenamefont {Lucero}, \citenamefont {Megrant},
  \citenamefont {Mutus}, \citenamefont {Neeley}, \citenamefont {Neill},
  \citenamefont {Quintana}, \citenamefont {Sank}, \citenamefont {Vainsencher},
  \citenamefont {Wenner}, \citenamefont {White}, \citenamefont {Coveney},
  \citenamefont {Love}, \citenamefont {Neven}, \citenamefont {Aspuru-Guzik},\
  and\ \citenamefont {Martinis}}]{PhysRevX.6.031007}%
  \BibitemOpen
  \bibfield  {author} {\bibinfo {author} {\bibfnamefont {P.~J.~J.}\
  \bibnamefont {O'Malley}}, \bibinfo {author} {\bibfnamefont {R.}~\bibnamefont
  {Babbush}}, \bibinfo {author} {\bibfnamefont {I.~D.}\ \bibnamefont
  {Kivlichan}}, \bibinfo {author} {\bibfnamefont {J.}~\bibnamefont {Romero}},
  \bibinfo {author} {\bibfnamefont {J.~R.}\ \bibnamefont {McClean}}, \bibinfo
  {author} {\bibfnamefont {R.}~\bibnamefont {Barends}}, \bibinfo {author}
  {\bibfnamefont {J.}~\bibnamefont {Kelly}}, \bibinfo {author} {\bibfnamefont
  {P.}~\bibnamefont {Roushan}}, \bibinfo {author} {\bibfnamefont
  {A.}~\bibnamefont {Tranter}}, \bibinfo {author} {\bibfnamefont
  {N.}~\bibnamefont {Ding}}, \bibinfo {author} {\bibfnamefont {B.}~\bibnamefont
  {Campbell}}, \bibinfo {author} {\bibfnamefont {Y.}~\bibnamefont {Chen}},
  \bibinfo {author} {\bibfnamefont {Z.}~\bibnamefont {Chen}}, \bibinfo {author}
  {\bibfnamefont {B.}~\bibnamefont {Chiaro}}, \bibinfo {author} {\bibfnamefont
  {A.}~\bibnamefont {Dunsworth}}, \bibinfo {author} {\bibfnamefont {A.~G.}\
  \bibnamefont {Fowler}}, \bibinfo {author} {\bibfnamefont {E.}~\bibnamefont
  {Jeffrey}}, \bibinfo {author} {\bibfnamefont {E.}~\bibnamefont {Lucero}},
  \bibinfo {author} {\bibfnamefont {A.}~\bibnamefont {Megrant}}, \bibinfo
  {author} {\bibfnamefont {J.~Y.}\ \bibnamefont {Mutus}}, \bibinfo {author}
  {\bibfnamefont {M.}~\bibnamefont {Neeley}}, \bibinfo {author} {\bibfnamefont
  {C.}~\bibnamefont {Neill}}, \bibinfo {author} {\bibfnamefont
  {C.}~\bibnamefont {Quintana}}, \bibinfo {author} {\bibfnamefont
  {D.}~\bibnamefont {Sank}}, \bibinfo {author} {\bibfnamefont {A.}~\bibnamefont
  {Vainsencher}}, \bibinfo {author} {\bibfnamefont {J.}~\bibnamefont {Wenner}},
  \bibinfo {author} {\bibfnamefont {T.~C.}\ \bibnamefont {White}}, \bibinfo
  {author} {\bibfnamefont {P.~V.}\ \bibnamefont {Coveney}}, \bibinfo {author}
  {\bibfnamefont {P.~J.}\ \bibnamefont {Love}}, \bibinfo {author}
  {\bibfnamefont {H.}~\bibnamefont {Neven}}, \bibinfo {author} {\bibfnamefont
  {A.}~\bibnamefont {Aspuru-Guzik}}, \ and\ \bibinfo {author} {\bibfnamefont
  {J.~M.}\ \bibnamefont {Martinis}},\ }\href {\doibase
  10.1103/PhysRevX.6.031007} {\bibfield  {journal} {\bibinfo  {journal} {Phys.
  Rev. X}\ }\textbf {\bibinfo {volume} {6}},\ \bibinfo {pages} {031007}
  (\bibinfo {year} {2016})}\BibitemShut {NoStop}%
\bibitem [{\citenamefont {Kandala}\ \emph {et~al.}(2017)\citenamefont
  {Kandala}, \citenamefont {Mezzacapo}, \citenamefont {Temme}, \citenamefont
  {Takita}, \citenamefont {Brink}, \citenamefont {Chow},\ and\ \citenamefont
  {Gambetta}}]{kandala2017hardware}%
  \BibitemOpen
  \bibfield  {author} {\bibinfo {author} {\bibfnamefont {A.}~\bibnamefont
  {Kandala}}, \bibinfo {author} {\bibfnamefont {A.}~\bibnamefont {Mezzacapo}},
  \bibinfo {author} {\bibfnamefont {K.}~\bibnamefont {Temme}}, \bibinfo
  {author} {\bibfnamefont {M.}~\bibnamefont {Takita}}, \bibinfo {author}
  {\bibfnamefont {M.}~\bibnamefont {Brink}}, \bibinfo {author} {\bibfnamefont
  {J.~M.}\ \bibnamefont {Chow}}, \ and\ \bibinfo {author} {\bibfnamefont
  {J.~M.}\ \bibnamefont {Gambetta}},\ }\href@noop {} {\bibfield  {journal}
  {\bibinfo  {journal} {Nature}\ }\textbf {\bibinfo {volume} {549}},\ \bibinfo
  {pages} {242} (\bibinfo {year} {2017})}\BibitemShut {NoStop}%
\bibitem [{\citenamefont {McClean}\ \emph {et~al.}(2016)\citenamefont
  {McClean}, \citenamefont {Romero}, \citenamefont {Babbush},\ and\
  \citenamefont {Aspuru-Guzik}}]{VQETheoryNJP}%
  \BibitemOpen
  \bibfield  {author} {\bibinfo {author} {\bibfnamefont {J.~R.}\ \bibnamefont
  {McClean}}, \bibinfo {author} {\bibfnamefont {J.}~\bibnamefont {Romero}},
  \bibinfo {author} {\bibfnamefont {R.}~\bibnamefont {Babbush}}, \ and\
  \bibinfo {author} {\bibfnamefont {A.}~\bibnamefont {Aspuru-Guzik}},\ }\href
  {http://stacks.iop.org/1367-2630/18/i=2/a=023023} {\bibfield  {journal}
  {\bibinfo  {journal} {New Journal of Physics}\ }\textbf {\bibinfo {volume}
  {18}},\ \bibinfo {pages} {023023} (\bibinfo {year} {2016})}\BibitemShut
  {NoStop}%
\bibitem [{\citenamefont {Wecker}\ \emph {et~al.}(2015)\citenamefont {Wecker},
  \citenamefont {Hastings},\ and\ \citenamefont {Troyer}}]{PhysRevA.92.042303}%
  \BibitemOpen
  \bibfield  {author} {\bibinfo {author} {\bibfnamefont {D.}~\bibnamefont
  {Wecker}}, \bibinfo {author} {\bibfnamefont {M.~B.}\ \bibnamefont
  {Hastings}}, \ and\ \bibinfo {author} {\bibfnamefont {M.}~\bibnamefont
  {Troyer}},\ }\href {\doibase 10.1103/PhysRevA.92.042303} {\bibfield
  {journal} {\bibinfo  {journal} {Phys. Rev. A}\ }\textbf {\bibinfo {volume}
  {92}},\ \bibinfo {pages} {042303} (\bibinfo {year} {2015})}\BibitemShut
  {NoStop}%
\bibitem [{\citenamefont {McClean}\ \emph {et~al.}(2017)\citenamefont
  {McClean}, \citenamefont {Kimchi-Schwartz}, \citenamefont {Carter},\ and\
  \citenamefont {de~Jong}}]{PhysRevA.95.042308}%
  \BibitemOpen
  \bibfield  {author} {\bibinfo {author} {\bibfnamefont {J.~R.}\ \bibnamefont
  {McClean}}, \bibinfo {author} {\bibfnamefont {M.~E.}\ \bibnamefont
  {Kimchi-Schwartz}}, \bibinfo {author} {\bibfnamefont {J.}~\bibnamefont
  {Carter}}, \ and\ \bibinfo {author} {\bibfnamefont {W.~A.}\ \bibnamefont
  {de~Jong}},\ }\href {\doibase 10.1103/PhysRevA.95.042308} {\bibfield
  {journal} {\bibinfo  {journal} {Phys. Rev. A}\ }\textbf {\bibinfo {volume}
  {95}},\ \bibinfo {pages} {042308} (\bibinfo {year} {2017})}\BibitemShut
  {NoStop}%
\bibitem [{\citenamefont {Colless}\ \emph {et~al.}(2018)\citenamefont
  {Colless}, \citenamefont {Ramasesh}, \citenamefont {Dahlen}, \citenamefont
  {Blok}, \citenamefont {Kimchi-Schwartz}, \citenamefont {McClean},
  \citenamefont {Carter}, \citenamefont {de~Jong},\ and\ \citenamefont
  {Siddiqi}}]{PhysRevX.8.011021}%
  \BibitemOpen
  \bibfield  {author} {\bibinfo {author} {\bibfnamefont {J.~I.}\ \bibnamefont
  {Colless}}, \bibinfo {author} {\bibfnamefont {V.~V.}\ \bibnamefont
  {Ramasesh}}, \bibinfo {author} {\bibfnamefont {D.}~\bibnamefont {Dahlen}},
  \bibinfo {author} {\bibfnamefont {M.~S.}\ \bibnamefont {Blok}}, \bibinfo
  {author} {\bibfnamefont {M.~E.}\ \bibnamefont {Kimchi-Schwartz}}, \bibinfo
  {author} {\bibfnamefont {J.~R.}\ \bibnamefont {McClean}}, \bibinfo {author}
  {\bibfnamefont {J.}~\bibnamefont {Carter}}, \bibinfo {author} {\bibfnamefont
  {W.~A.}\ \bibnamefont {de~Jong}}, \ and\ \bibinfo {author} {\bibfnamefont
  {I.}~\bibnamefont {Siddiqi}},\ }\href {\doibase 10.1103/PhysRevX.8.011021}
  {\bibfield  {journal} {\bibinfo  {journal} {Phys. Rev. X}\ }\textbf {\bibinfo
  {volume} {8}},\ \bibinfo {pages} {011021} (\bibinfo {year}
  {2018})}\BibitemShut {NoStop}%
\bibitem [{\citenamefont {McArdle}\ \emph
  {et~al.}(2018{\natexlab{b}})\citenamefont {McArdle}, \citenamefont {Endo},
  \citenamefont {Jones}, \citenamefont {Li}, \citenamefont {Benjamin},\ and\
  \citenamefont {Yuan}}]{mcardle2018variational}%
  \BibitemOpen
  \bibfield  {author} {\bibinfo {author} {\bibfnamefont {S.}~\bibnamefont
  {McArdle}}, \bibinfo {author} {\bibfnamefont {S.}~\bibnamefont {Endo}},
  \bibinfo {author} {\bibfnamefont {T.}~\bibnamefont {Jones}}, \bibinfo
  {author} {\bibfnamefont {Y.}~\bibnamefont {Li}}, \bibinfo {author}
  {\bibfnamefont {S.}~\bibnamefont {Benjamin}}, \ and\ \bibinfo {author}
  {\bibfnamefont {X.}~\bibnamefont {Yuan}},\ }\href@noop {} {\bibfield
  {journal} {\bibinfo  {journal} {arXiv preprint arXiv:1804.03023}\ } (\bibinfo
  {year} {2018}{\natexlab{b}})}\BibitemShut {NoStop}%
\bibitem [{\citenamefont {Endo}\ \emph
  {et~al.}(2018{\natexlab{b}})\citenamefont {Endo}, \citenamefont {Jones},
  \citenamefont {McArdle}, \citenamefont {Yuan},\ and\ \citenamefont
  {Benjamin}}]{endo2018discovering}%
  \BibitemOpen
  \bibfield  {author} {\bibinfo {author} {\bibfnamefont {S.}~\bibnamefont
  {Endo}}, \bibinfo {author} {\bibfnamefont {T.}~\bibnamefont {Jones}},
  \bibinfo {author} {\bibfnamefont {S.}~\bibnamefont {McArdle}}, \bibinfo
  {author} {\bibfnamefont {X.}~\bibnamefont {Yuan}}, \ and\ \bibinfo {author}
  {\bibfnamefont {S.}~\bibnamefont {Benjamin}},\ }\href@noop {} {\bibfield
  {journal} {\bibinfo  {journal} {arXiv preprint arXiv:1806.05707}\ } (\bibinfo
  {year} {2018}{\natexlab{b}})}\BibitemShut {NoStop}%
\end{thebibliography}%

\end{document}